\documentclass[twocolumn,preprintnumbers,amsmath,amssymb,superscriptaddress]{revtex4-1}
\usepackage{graphicx}% Include figure files
\usepackage{dcolumn}% Align table columns on decimal point
\usepackage{bm}% bold math

\begin{document}

%\preprint{APS/123-QED}

\title{Observation of a Slater-type metal-to-insulator transition in Sr$_2$IrO$_4$ from time-resolved photo-carrier dynamics}% Force line breaks with \\

\author{D. Hsieh}
\affiliation{Department of Physics, Massachusetts Institute of
Technology, Cambridge, MA 02139, USA}
\author{F. Mahmood}
\affiliation{Department of Physics, Massachusetts Institute of
Technology, Cambridge, MA 02139, USA}
\author{D. H. Torchinsky}
\affiliation{Department of Physics, Massachusetts Institute of
Technology, Cambridge, MA 02139, USA}
\author{G. Cao}
\affiliation{Center for Advanced Materials, University of Kentucky, Lexington, Kentucky 40506, USA}
\affiliation{Department of Physics and Astronomy, University of Kentucky, Lexington, Kentucky 40506, USA}
\author{N. Gedik}
\affiliation{Department of Physics, Massachusetts Institute of
Technology, Cambridge, MA 02139, USA}

\date{\today}% It is always \today, today,


\begin{abstract}
We perform a time-resolved optical study of Sr$_2$IrO$_4$ to understand the influence of magnetic ordering on the low energy electronic structure of a strongly spin-orbit coupled $J_{eff}$=1/2 Mott insulator. By studying the recovery dynamics of photo-carriers excited across the Mott gap, we find that upon cooling through the N\'{e}el temperature $T_N$ the system evolves continuously from a metal-like phase with fast ($\sim$50 fs) and excitation density independent relaxation dynamics to a gapped phase characterized by slower ($\sim$500 fs) excitation density dependent bimolecular recombination dynamics. The development of the insulating gap is accompanied by a transfer of in-gap spectral weight to energies far in excess of the gap and occurs over an unusually broad temperature window, which suggests Sr$_2$IrO$_4$ to be a Slater- rather than Mott-Hubbard type insulator and naturally explains the absence of anomalies at $T_N$ in transport and thermodynamic measurements.
\end{abstract}

\maketitle

Iridium oxides are unique 5$d$ electronic systems in which spin-orbit coupling, electronic bandwidth ($W$) and on-site Coulomb interactions ($U$) occur on comparable energy scales. Their interplay can stabilize a novel $J_{eff}$=1/2 Mott insulating state in which a correlation gap is opened by only moderate Coulomb interactions owing to a spin-orbit coupling induced band narrowing \cite{Kim_PRL}. Depending on the underlying lattice, this insulating state is predicted to realize a variety of exotic quantum phases including antiferromagnetic $J_{eff}$=1/2 Mott insulators on the perovskite lattice \cite{Cao,Kim_PRL,Okabe}, correlated topological insulators and semimetals on the pyrochlore lattice \cite{Pesin,Yang,Wan} and topological spin liquids on the hyper-kagome \cite{Lawler} and honeycomb lattices \cite{You}. Even more tantalizing possibilities are predicted to occur upon chemically doping these systems, ranging from high-$T_c$ superconductivity \cite{Fa,Martins} to spin-triplet superconductivity \cite{You}.

Intensive research has been conducted on the layered perovskite iridate Sr$_2$IrO$_4$ owing to its structural and electronic similarities to undoped high-$T_c$ cuprates such as La$_2$CuO$_4$ \cite{Kim_PRL,Fa}. The ground state electronic structure of Sr$_2$IrO$_4$ consists of a completely filled band with total angular momentum $J_{eff}=3/2$ and a narrow half-filled $J_{eff}=1/2$ band near the Fermi level $E_F$ \cite{Kim_PRL,Kim_Science,Watanabe}. The latter is split into an upper Hubbard band (UHB) and lower Hubbard band (LHB) due to on-site Coulomb interactions and exhibits antiferromagnetic ordering of the effective $J_{eff}=1/2$ moments below a N\'{e}el temperature $T_N$ = 240 K \cite{Cao} analogous to La$_2$CuO$_4$. Although this insulating ground state has been established by angle-resolved photoemission spectroscopy \cite{Kim_PRL} and resonant x-ray scattering \cite{Kim_Science} measurements, whether Sr$_2$IrO$_4$ is a Mott-type ($U \gg W$) insulating phase typical of 3$d$ transition metal oxides or a Slater-type ($U \approx W$) insulating phase is experimentally unknown and remains a topic of active theoretical debate \cite{Jackeli,Martins,Arita}. Whereas a Mott-Hubbard type metal-to-insulator transition (MIT) is discontinuous and occurs at temperatures greater or equal to $T_N$, a Slater-type MIT is continuous and occurs exactly at $T_N$ \cite{Gebhard}. Therefore the relevant experimental question is whether heating above $T_N$ brings Sr$_2$IrO$_4$ into a paramagnetic insulating phase or into a paramagnetic metallic phase and what the order of the MIT is.

Owing to an absence of clear anomalies at $T_N$ in transport \cite{Cao,Kini,Ge,Chikara}, thermodynamic \cite{Chikara} and optical conductivity data \cite{Moon}, there have been conflicting interpretations about how the insulating gap behaves across $T_N$. In this Letter we use time-resolved optical spectroscopy, which is highly sensitive to the existence of energy gaps \cite{Gedik,Chia}, to study the temperature evolution of the electronic structure of Sr$_2$IrO$_4$. Taking advantage of qualitatively distinct relaxation dynamics of photo-excited carriers exhibited by gapped and gapless systems, we find a clear change in the ultrafast dynamics across $T_N$ indicating a gap opening concomitant with antiferromagnetic order. Analysis of the long time relaxation dynamics further reveals that in-gap spectral weight is continuously transferred to high energies through $T_N$, consistent with a phase transition from low-temperature Slater-type insulator to high-temperature paramagnetic metal.

%We are in weak excitation limit where we do not drive transition to a metal.

\begin{figure}
\includegraphics[scale=0.34,clip=true, viewport=0.0in 0in 10.8in 7.3in]{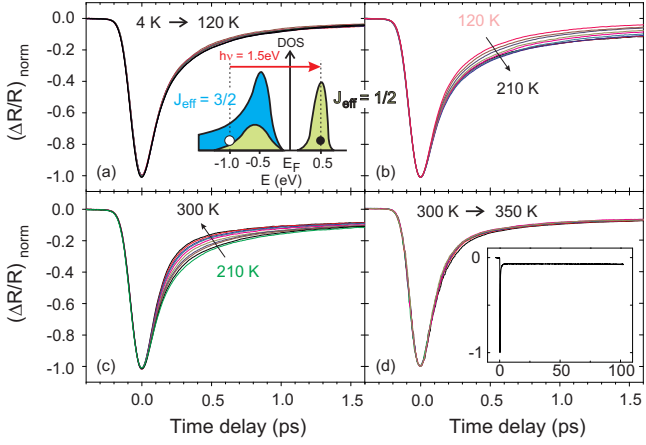}% Here is how to import EPS art
\caption{\label{fig:1} Normalized time-resolved reflectivity traces ($\Delta$R/R)$_{norm}$ of Sr$_2$IrO$_4$ collected in the temperature ranges (a) 4 K to 120 K, (b) 120 K to 210 K, (c) 210 K to 300 K and (d) 300 K to 350 K. Curves are collected in 10 K intervals with a pump fluence of 15.4 $\mu$J/cm$^2$. Inset of panel (a) shows a schematic of the low energy electronic density of states (DOS) based on calculations \cite{Watanabe}. The red arrow denotes the optical transition being excited by the pump pulse. Inset of panel (d) shows the $T$ = 300 K trace out to 100 ps time delay.}
\end{figure}

In our experiment we used a Ti:Sapphire oscillator producing laser pulses with center wavelength 795 nm ($h\nu = 1.56$ eV) and near 80 fs duration. The fluence of the pump pulse, which excites electrons from the occupied $J_{eff}=3/2$ and lower Hubbard $J_{eff}=1/2$ bands to the upper Hubbard $J_{eff}=1/2$ band [inset Fig.~\ref{fig:1}(a)], was varied with neutral density filters to tune the photo-excited carrier density while the probe fluence was maintained at 4 $\mu$J/cm$^2$. Both beams were focused to a 70 $\mu$m FWHM spot on the (001) cleaved face of the sample. The 80 MHz repetition rate was reduced to 1.6 MHz with a pulse picker to eliminate steady state heating of the sample. Use of a double-modulation scheme \cite{Gedik} provided sensitivity to the fractional change of reflectivity on the order $\Delta$R/R $\sim 10^{-7}$. Single crystals of Sr$_2$IrO$_4$ were grown using a self-flux technique and magnetization measurements show a magnetic ordering temperature at $T_N$ = 240 K \cite{Cao,Chikara}.

Figure 1 shows typical time-resolved reflectivity transients measured over a range of initial temperatures spanning 4 K to 350 K, which have all been normalized to their negative peak values in order to emphasize the recovery dynamics. Following the pump excitation all $\Delta$R/R traces exhibit a rapid negative spike, which indicates a decrease in reflectivity. Within approximately 1 ps, the reflectivity recovers to a small negative offset that persists beyond 100 ps [inset Fig.~\ref{fig:1}(d)]. By performing a temperature dependence study we find that the initial $\sim$1 ps recovery can actually be decomposed into two separate components as follows. Figure~\ref{fig:1}(a) shows that between 4 K and 120 K there is no discernible change in the recovery dynamics. Upon heating from 120 K to 210 K [Fig.~\ref{fig:1}(b)], we find no temperature dependence within the first $\sim$100 fs but observe a clear slowing down of recovery dynamics after this time. Further heating of the sample from 210 K to 300 K again causes no change in the first recovery component but causes the second component to become faster [Fig.~\ref{fig:1}(c)]. All temperature dependence then completely shuts off above 300 K [Fig.~\ref{fig:1}(d)]. To understand the physical processes underlying these trends, we perform detailed fits to the $\Delta$R/R traces.

\begin{figure}
\includegraphics[scale=0.33,clip=true, viewport=0.0in 0in 10.8in 8.3in]{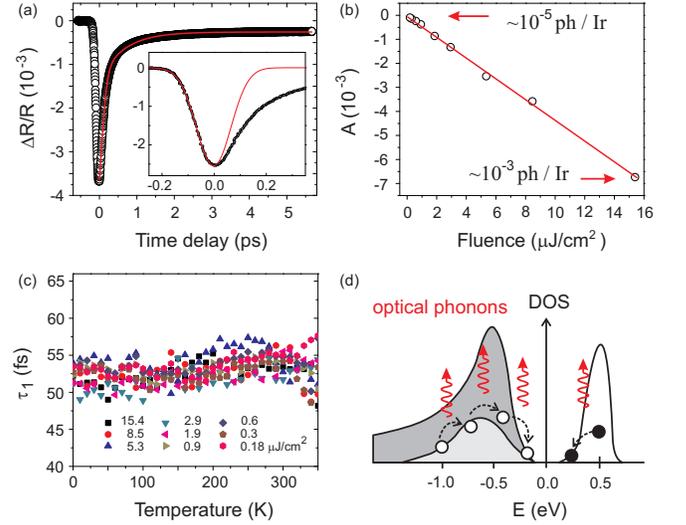}% Here is how to import EPS art
\caption{\label{fig:2} (a) A typical un-normalized time-resolved reflectivity transient ($\Delta$R/R) of Sr$_2$IrO$_4$ overlaid with a bi-exponential fit described in the text (red curve). Inset shows the Gaussian instrument time resolution (red curve) superimposed on 300 K data where decay dynamics are dominated by the fast ($\tau_1$) component [Fig.~\ref{fig:3}(a)]. This shows that the fast initial decay is not resolution limited. (b) The fitted amplitude (A) of the fast exponential decay component measured over a range of pump fluences corresponding to $\sim$ 10$^{-5}$ to $\sim$ 10$^{-3}$ pump photons per iridium atom. The red line is a linear fit. (c) Temperature dependence of the fitted decay time ($\tau_1$) of the fast exponential decay component measured over a range of pump fluences. (d) Schematic showing that the initial fast decay process is governed by energy relaxation of photo-excited electrons and holes towards the band edges via optical phonon emission.}
\end{figure}

The reflectivity transients at all temperatures are well described by a bi-exponential function Ae$^{t/\tau_1}$+Be$^{t/\tau_2}$+C convolved with our Gaussian instrument resolution as demonstrated in Figure~\ref{fig:2}(a). We begin by trying to understand the physical origin of the first decay component, whose decay time $\tau_1$ is within our ability to resolve [inset Fig.~\ref{fig:2}(a)]. To investigate whether this fast decay arises from thermalization of photo-excited carriers via carrier-carrier scattering, whose rate should depend on the number of photo-excited carriers \cite{Hohlfeld}, we measured $\tau_1$ over a range of pump fluences spanning $\sim$10$^{-5}$ to $\sim$10$^{-3}$ photons per iridium site \cite{EPAPS}. A linear behavior of the component amplitude (A) over this fluence range [Fig.~\ref{fig:2}(b)] shows that the number of photo-excited carriers is indeed proportional to the pump fluence. Figure~\ref{fig:2}(c) shows that $\tau_ 1 \approx$ 50 fs exhibits no discernible fluence nor temperature dependence between 4 K and 350 K. This implies that the initial decay is not caused by photo-carrier thermalization but more likely by photo-carrier cooling, and that this cooling is not mediated by thermally occupied phonons. The participation of magnons is also negligible because $\tau_1$ is insensitive to $T_N$. Given that the fast timescale of $\tau_1$ is consistent with typical optical phonon mediated cooling processes \cite{Wang} and that the Debye temperature of Sr$_2$IrO$_4$ \cite{Kini} far exceeds our measurement temperatures, we conclude that the initial fast recovery of $\Delta$R/R is due to the cooling of photo-excited carriers via generation of hot optical phonons [Fig.~\ref{fig:2}(d)].

\begin{figure}
\includegraphics[scale=0.38,clip=true, viewport=0.0in 0in 10.8in 8.5in]{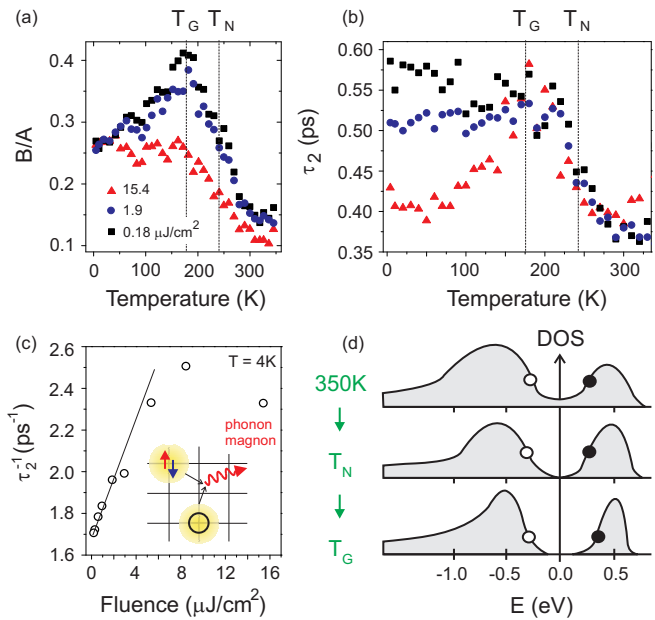}% Here is how to import EPS art
\caption{\label{fig:3} Temperature dependence of the (a) amplitude ratio of the slow to fast decay components (B/A) and (b) the decay time of the slower component ($\tau_2$) measured using three different fluences 15.4 $\mu$J/cm$^2$, 1.9 $\mu$J/cm$^2$ and 0.18 $\mu$J/cm$^2$. (c) Fluence dependence of the slower decay rate measured at 4 K. Straight line is a guide to the eye showing a linear dependence at low fluences. A schematic of the bimolecular relaxation process involving the annihilation of photo-excited empty and doubly-occupied sites via emission of optical phonons or magnons is shown in the inset. (d) Schematic showing the temperature evolution of the low energy electronic density of states (DOS) in Sr$_2$IrO$_4$ based on our data.}
\end{figure}

The initial fast recovery of $\Delta$R/R is followed by a second slower decay component, whose relative amplitude (B/A) exhibits an upturn upon cooling through $T_N$ and then ceases to grow further below a temperature $T_G = 175$ K [Fig.~\ref{fig:3}(a)]. This indicates that at temperatures far above $T_N$ the electronic energy relaxation occurs predominantly through the lone process of optical phonon generation described in Figure~\ref{fig:2} and that a separate relaxation mechanism grows near $T_N$ before saturating below $T_G$. The decay time $\tau_2$ shows a similar upturn upon cooling through $T_N$ from around 0.36 ps at 350 K to 0.55 ps at $T_G$ and then exhibits a marked change in temperature dependence below $T_G$. A sharp rise in relaxation time at $T_N$ typically signifies the development of an energy gap \cite{Chia,Gedik}, where a depletion of states around $E_F$ greatly reduces the efficiency of photo-carrier relaxation from above to below $E_F$. The hallmark of a fully formed energy gap is that photo-excited occupied states above $E_F$ and empty states below $E_F$ can only combine in a pairwise fashion. Therefore, unlike in a metal, the relaxation rate should be proportional to the density of photo-excited carriers \cite{Rothwarf}. We find that a clear fluence dependence of $\tau_2$ develops only below $T_G$ [Fig.~\ref{fig:3}(b)] and that the low temperature relaxation rate $\tau_2^{-1}$ indeed increases linearly with fluence below 5 $\mu$J/cm$^2$ [Fig.~\ref{fig:3}(c)]. This shows that below $T_G$, $\tau_2$ represents the timescale for recombination of photo-excited singly- and doubly-occupied sites across the correlation gap [inset Fig.~\ref{fig:3}(c)]. This ultrafast sub-ps recombination time in Sr$_2$IrO$_4$ is comparable to that observed in the 2D antiferromagnetic cuprate Mott insulators \cite{Okamoto}, which has been attributed to additional recombination channels involving hot magnon generation. Altogether these observations allow us to conclude that at temperatures far above $T_N$ the system behaves like a metal. Near $T_N$ the density of states around $E_F$ is continually depleted upon cooling, which is consistent with optical conductivity measurements \cite{Moon}, and the gap between the UHB and LHB is fully developed below $T_G$.

%Note flattening below 100 K. Speculate about why cooling gets faster as you cool.

\begin{figure}
\includegraphics[scale=0.31,clip=true, viewport=0.0in 0in 10.8in 4.5in]{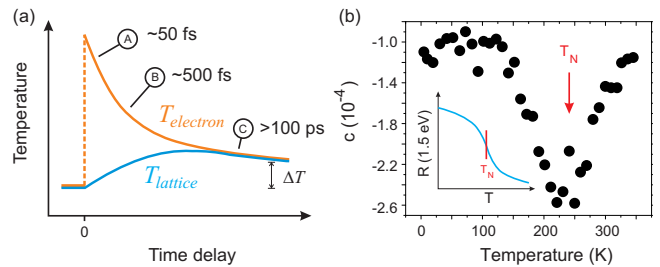}% Here is how to import EPS art
\caption{\label{fig:4} (a) Diagram illustrating the temporal equilibration between the electronic and lattice sub-systems following a pump pulse. The slowest process (C) is the cooling of the equilibrated system via diffusion of hot carriers or hot phonons away from the measured sample spot. (b) Temperature dependence of the offset (C) measured using a fluence of 15.4 $\mu$J/cm$^2$. A clear minimum in C is observed at $T_N$ corresponding to a rapid rise in reflectivity at 1.5 eV at $T_N$ (see inset schematic).}
\end{figure}

The occurrence of the metal-to-insulator transition at $T_N$ in Sr$_2$IrO$_4$ (Fig.~\ref{fig:3}) distinguishes it from archetypal Mott insulators such as MnO where the insulating gap persists even in the absence of long-range magnetic order \cite{Imada}. The continuous nature of the MIT and its development over a broad (0.7 $\lesssim T/T_N \lesssim$ 1.4) temperature window further precludes a Mott-Hubbard description, which predicts sharp first-order MIT's like in V$_2$O$_3$ \cite{Imada}. We rule out the possibility of a disorder broadened $T_{MIT}$ in our samples based on their sharp magnetic susceptibility curves \cite{Chikara,Cao}. Although these results point towards a second-order Slater-type MIT, Sr$_2$IrO$_4$ does not conform to a weakly correlated ($U \ll W$) spin-density wave description, which are driven purely by Fermi surface nesting and are typically only partially gapped below $T_N$. Unlike a conventional spin-density wave system, magnetic ordering in Sr$_2$IrO$_4$ does not change the size of the unit cell \cite{Kim_Science} and local moment fluctuations exist well above $T_N$ according to both magnetic susceptibility \cite{Chikara,Cao} and magnetic diffuse x-ray scattering measurements \cite{Fujiyama}. While these magnetic signatures provide evidence for correlation physics at play, a direct electronic distinction between a correlation driven MIT and a Fermi surface nesting driven MIT is whether or not, respectively, spectral weight is transferred from low (near $E_F$) energies to high (of order $U$) energies upon traversing the MIT \cite{Kotliar}.

To determine whether the in-gap spectral weight lost at $T_N$ is being transferred to upper and lower Hubbard bands, we investigate the temperature dependence of the optical reflectivity at the energy scale of the LHB and UHB splitting [inset Fig.~\ref{fig:1}(a)] by studying the small negative $\Delta$R/R offset term C. The two exponential relaxation processes have been identified as intra-band cooling and recombination processes that both transfer energy from the electronic to lattice sub-systems, which brings them into thermal equilibrium within the first few ps [Fig.~\ref{fig:4}(a)]. Subsequent cooling of this heated sample spot back to the initial temperature takes place through the diffusion of hot carriers or hot phonons away from the laser illuminated area, which is estimated to well exceed 100 ps \cite{EPAPS} based on thermal conductivity and heat capacity data for Sr$_2$IrO$_4$ \cite{Kini}. Therefore to a very good approximation the offset C is a measure of the fractional change in reflectivity at 1.5 eV due to a small temperature change $\Delta T$ [Fig.~\ref{fig:4}(a)], namely C $=\frac{R(T+\Delta T)-R(T)}{R(T)}$. We estimate that at high pump fluences $\Delta T$ is of order 1 K because C is of order 10$^{-4}$ and the fractional decrease in optical conductivity at 1.5 eV is estimated to be 2\% from 10 K to 500 K \cite{Moon}.

Figure~\ref{fig:4}(b) shows that the temperature dependence of C exhibits a broad negative peak with an extremum exactly at $T_N$, which indicates that a small temperature rise causes the largest decrease of the 1.5 eV reflectivity exactly at $T_N$. From this we can infer a broad temperature window (0.6 $\lesssim T/T_N \lesssim$ 1.4) within which R(1.5 eV) starts to rise sharply with cooling, with the greatest slope occuring at $T_N$ [inset Fig.~\ref{fig:4}(b)]. The fact that this temperature window largely coincides with that over which $\tau_2$ increases most drastically [Fig.~\ref{fig:3}(b)] shows that spectral weight is being transferred from low in-gap energies ($\sim$0.1 eV \cite{Watanabe}) to energies (1.5 eV) far exceeding it.

Our results taken altogether reveal that unlike La$_2$CuO$_4$, which remains a Mott insulator far above its magnetic ordering temperature, Sr$_2$IrO$_4$ undergoes a metal-to-insulator transition across $T_N$. This suggests that Sr$_2$IrO$_4$ is more accurately described as an intermediate coupling ($U \approx W$) Slater-type insulator with key Mott-Hubbard characteristics. Our observation that this metal-to-insulator transition takes place over a wide temperature window compared to $T_N$ naturally explains the lack of sharp anomalies at $T_N$ in transport, thermodynamic and optical conductivity data. We argue that these may be signatures of a rare example of a temperature controlled continuous metal-to-insulator transition in a quasi-two dimensional system hitherto unobserved in any $d$-electron material \cite{Imada}, and may be also applicable to the wider class of $J_{eff}$=1/2 Mott insulating perovskite, honeycomb, hyperkagome and pyrochlore iridates. Moreover, transport data on 5$d$ Os oxides suggestive of continuous metal-to-insulators transitions \cite{Mandrus} show that the confluence of strong spin-orbit coupling and on-site Coulomb interactions is a general playground for unconventional metal-to-insulator transitions.

%In every section put some emphasis on why pump-probe is much more sensitive. Can STM look at spectral weight at 1.5eV?

\vspace{0.5cm}

\noindent \textbf{Acknowledgements.}

We thank F. Wang, T. Senthil, A. Vishwanath, S. Kehrein, K. Michaeli, R. Flint, S. Drapcho and Y. Wang for useful discussions.

N. G. acknowledges support by Army Research Office grant number W911NF-11-1-0331. C. G. acknowledges support through NSF grants DMR-0856234 and EPS-0814194.

%D.H. is supported through a Pappalardo Postdoctoral Fellowship.

\end{document}